\documentclass{article}
\usepackage{jheppub}
\usepackage{graphicx}
\usepackage{dcolumn}
\usepackage{bm}
\usepackage{physics}
\usepackage{amsmath,amsfonts,amssymb}
\usepackage{subcaption}
\usepackage{setspace}
\usepackage{color}
\usepackage{tikz}
\usepackage{float}
\usepackage{hyperref}
\usepackage{parskip}
\usepackage{cleveref}
\usepackage[normalem]{ulem}
\usepackage{tikz}
\usetikzlibrary{shadings, calc}

\definecolor{gold}{RGB}{255, 215, 0}
\definecolor{darkgold}{RGB}{184, 134, 11}

\allowdisplaybreaks
\title{Generalized Brick Wall Method for Stationary Axisymmetric Spacetimes}

\author[a]{Chandra Prakash}

\affiliation[a]{Department of Physics, Indian Institute of Technology, Guwahati, 
	Assam, India}

\emailAdd{chandra.pp@alumni.iitg.ac.in}

\abstract{The microscopic origin of black hole entropy remains one of the central puzzles in quantum gravity. In this work, we investigate the statistical entropy of scalar fields propagating in stationary axisymmetric spacetimes using the thin-film modification of the 't Hooft brick wall method. We derive a generalized expression for the free energy of both superradiant and non-superradiant modes, expressed explicitly in terms of generic metric components. This unified formalism allows for a systematic evaluation of entropy across a diverse class of black holes without re-deriving the wave equation for each specific case. We validate our approach by recovering the Bekenstein-Hawking area law for the standard Kerr black hole. Subsequently, we extend the analysis to the Kerr-Newman-AdS geometry and, finally, to the novel case of a Kerr-Newman-AdS black hole surrounded by quintessence and a cloud of strings. Our results confirm that the area law holds even in the presence of these complex background matter fields, provided the cutoff parameter is appropriately renormalized.}

\makeatletter
\def\@fpheader{\ }
\makeatother
\date{} 
\begin{document}
	
	\maketitle
\section{Introduction}

Since the seminal works of Bekenstein and Hawking~\cite{bekenstein1973black, hawking1975particle, bardeen1973four}, the thermodynamic properties of black holes have occupied a central role in theoretical physics. The realization that black holes possess an entropy proportional to their horizon area, $S = A/4$, suggests a deep connection between gravitation, thermodynamics, and quantum mechanics~\cite{bekenstein1974generalized, jacobson1995thermodynamics, wald1993black}. However, identifying the microscopic degrees of freedom responsible for this entropy remains an open challenge~\cite{strominger1996microscopic, hooft1985quantum, srednicki1993entropy, hawking1976breakdown}.

A significant step towards understanding this statistical origin was taken by 't Hooft in his ``brick wall'' model~\cite{hooft1985quantum}. By quantizing a scalar field in the black hole background, 't Hooft demonstrated that the entropy arises from the density of states of quantum fields near the event horizon. While the original method required a large-distance cutoff to regularize the vacuum contribution, subsequent refinements (thin film model) have shown that the dominant contribution to the entropy comes strictly from a narrow region near the horizon.\cite{wen2001improved,liu2001improved,zhang2002entropy,sun2004improved,zhou2003entropy} Physically, the divergence of the local temperature at the horizon effectively decouples the radial integration; although the calculation is formally a volume integral, the overwhelming contribution from the near-horizon pole forces the result to scale with the area rather than the volume. This eliminates the need for an external boundary and focuses the calculation on the near-horizon geometry.

Despite its success in reproducing the area law, the thin-film model should be interpreted as a phenomenological effective description rather than a fundamental microscopic theory. The regularization relies on an ultraviolet cutoff $\epsilon$, introduced manually to render the entropy finite. Since the model does not derive this scale from first principles, it remains agnostic regarding the precise identity of the quantum microstates, whether they be strings, spin networks, or other fundamental degrees of freedom. Crucially, the model's consistency relies on the breakdown of semiclassical approximations at the horizon: if the local temperature did not diverge, the entropy would scale with volume rather than area. In this sense, the thin film serves as a mathematical manifestation of the unknown quantum gravity physics at the Planck scale.

\begin{figure}[H]
	\begin{tikzpicture}
		\def\rH{2.0}           
		\def\thickness{0.6}    
		\def\rOuter{\rH + \thickness} 
		\def\numParticles{400} 
		
		\foreach \i in {1,...,\numParticles} {
			\pgfmathsetmacro{\angle}{random(0,360)}
			\pgfmathsetmacro{\rad}{random()*\thickness + \rH} 
			\pgfmathsetmacro{\x}{\rad*cos(\angle)}
			\pgfmathsetmacro{\y}{\rad*sin(\angle)}
			
			\fill[orange!80!red, opacity=0.6] (\x, \y) circle (1.5pt);
		}
		
		\shade[ball color=black] (0,0) circle (\rH);
		\draw[color=white, opacity=0.1, line width=1pt] (0,0) circle (\rH);
		
		\foreach \i in {1,...,150} {
			\pgfmathsetmacro{\angle}{random(0,360)}
			\pgfmathsetmacro{\rad}{random()*\rH} 
			\pgfmathsetmacro{\x}{\rad*cos(\angle)}
			\pgfmathsetmacro{\y}{\rad*sin(\angle)}
			
			\fill[orange!90!yellow, opacity=0.7] (\x, \y) circle (1.5pt);
		}
		
		\shade[ball color=gold, opacity=0.15] (0,0) circle (\rOuter);
		\draw[darkgold, dotted, thin] (0,0) circle (\rOuter);
		
		\draw[<-] (0,0) -- ++(2.5, 2.5) node[right, align=left] {\textbf{Black Hole}\\(Interior)};
		
		\draw[<-, thick] (0, \rH + \thickness/2) -- ++(-3, 1) 
		node[left, align=right] {\textbf{Thin Film Region}\\(Entropy Source)};
		
		\draw[<-] (\rH + \thickness/2, 0) -- ++(1.5, -1.5) 
		node[right, align=left] {Quantum Degrees\\of Freedom ("Blobs")};
		
		\draw[|-|, thick, color=black] (0, -\rH) -- (0, -\rOuter) 
		node[midway, right, xshift=2pt] {$\delta$};
		
	\end{tikzpicture}
	\caption{Visualization of the "Thin Film" model. The entropy is dominated by quantum fluctuations (orange blobs) trapped in a thin shell of characteristic thickness $\delta$ near the event horizon. This localization explains the area proportionality of black hole entropy despite the extensive nature of the field modes.}
\end{figure}

In this paper, we employ the thin-film brick wall formalism to study the thermodynamics of rotating charged black holes in modified environments. This framework provides a transparent geometric interpretation of the entropy: by restricting the integration to a thin shell, the volume enclosed within the film naturally scales as the horizon area in the limit where the thickness approaches zero. Rather than solving the Klein-Gordon equation for specific cases individually, we first derive a \textit{generalized} expression for the free energy in terms of arbitrary metric components for stationary axisymmetric spacetimes. We pay particular attention to the regularization of divergent integrals, carefully separating the contributions from superradiant (SR) and non-superradiant (NSR) modes.

We apply this generalized framework to three increasingly complex scenarios. First, we revisit the standard Kerr black hole to verify the consistency of our formalism. Second, we extend the calculation to the Kerr-Newman-AdS spacetime. Finally, we investigate the thermodynamic properties of a Kerr-Newman-AdS black hole surrounded by quintessence and a cloud of strings. This latter metric is of particular interest as it interpolates between standard electro-vacuum solutions and those embedded in non-trivial cosmological matter distributions.

The paper is organized as follows: In Section II and Section III, we establish the Klein-Gordon equation in a general axisymmetric metric and derive the formal expressions for free energy. In Section IV, we apply this formalism to the Kerr, Kerr-Newman-AdS and Kerr-Newman-AdS with the quintessence and cloud of strings background. We conclude with a discussion of our results in Section V.

\section{Klein-Gordon Equation in General Axisymmetric Spacetimes}

In this section, we formulate the scalar field dynamics abstractly. While the explicit form of the Kerr-Newman line element is well known, we begin by considering a general stationary, axisymmetric metric of the form:
\begin{equation}
	ds^2 = g_{tt}dt^2 + g_{rr}dr^2 + g_{\theta\theta}d\theta^2 + g_{\phi\phi}d\phi^2 + 2g_{t\phi}dt d\phi
\end{equation}
This generalized approach allows us to isolate the specific metric dependencies of the superradiant and non-superradiant contributions to the free energy. The contravariant (inverse) metric components for the general line element are given by the matrix:
\begin{equation}\label{inv-metric}
	g^{\mu\nu} = 
	\begin{pmatrix}
		\frac{g_{\phi \phi }}{\mathcal{D}} & 0 & 0 & \frac{-g_{t\phi}}{\mathcal{D}} \\
		0 & \frac{1}{g_{rr}} & 0 & 0 \\
		0 & 0 & \frac{1}{g_{\theta \theta }} & 0 \\
		\frac{-g_{t\phi}}{\mathcal{D}} & 0 & 0 & \frac{g_{tt}}{\mathcal{D}} 
	\end{pmatrix}
\end{equation}
where $\mathcal{D} = g_{tt}g_{\phi\phi} - g_{t\phi}^2$.

\subsection{The Wave Equation}
The dynamics of a massless scalar field $\Phi$ are governed by the covariant Klein-Gordon equation:
\begin{equation}
	\frac{1}{\sqrt{-g}}\partial_\mu \left(\sqrt{-g}g^{\mu\nu}\partial_\nu \Phi \right) = 0
\end{equation}
Expanding the summation over indices $\mu, \nu \in \{t, r, \theta, \phi\}$ explicitly, we obtain:
\begin{multline}
	\frac{1}{\sqrt{-g}}\frac{\partial}{\partial t}\left(\sqrt{-g}g^{tt}\frac{\partial}{\partial t}\right) \Phi + \frac{1}{\sqrt{-g}}\frac{\partial}{\partial r}\left(\sqrt{-g}g^{rr}\frac{\partial}{\partial r}\right) \Phi + \frac{1}{\sqrt{-g}}\frac{\partial}{\partial \phi}\left(\sqrt{-g}g^{\phi \phi}\frac{\partial}{\partial \phi}\right) \Phi \\
	+ \frac{1}{\sqrt{-g}}\frac{\partial}{\partial \theta}\left(\sqrt{-g}g^{\theta\theta}\frac{\partial}{\partial \theta}\right) \Phi
	+ \frac{1}{\sqrt{-g}}\frac{\partial}{\partial t}\left(\sqrt{-g}g^{t \phi}\frac{\partial}{\partial \phi}\right) \Phi + \frac{1}{\sqrt{-g}}\frac{\partial}{\partial \phi}\left(\sqrt{-g}g^{\phi t}\frac{\partial}{\partial t}\right) \Phi = 0
\end{multline}
We now exploit the symmetries of the spacetime. Since the metric components are independent of time $t$ and the azimuthal angle $\phi$ (stationarity and axisymmetry), the determinant $\sqrt{-g}$ and the inverse metric components can be pulled out of the temporal and azimuthal derivatives. The equation simplifies to:
\begin{equation}
	g^{tt}\frac{\partial^2 \Phi}{\partial t^2} + \frac{1}{\sqrt{-g}}\frac{\partial}{\partial r}\left(\sqrt{-g}g^{rr}\frac{\partial \Phi}{\partial r}\right) + g^{\phi \phi}\frac{\partial^2 \Phi}{\partial \phi^2} + \frac{1}{\sqrt{-g}}\frac{\partial}{\partial \theta}\left(\sqrt{-g}g^{\theta\theta}\frac{\partial \Phi}{\partial \theta}\right) + 2g^{t \phi}\frac{\partial^2 \Phi}{\partial t\partial \phi} = 0
\end{equation}
To solve this, we employ the WKB approximation (or geometric optics limit). We assume a solution of the form:
\begin{equation}
	\Phi(t,r,\theta,\phi) = \exp\left[ -iEt + i m \phi + i R(r) + i S(\theta) \right]
\end{equation}
where $E$ is the energy, $m$ is the azimuthal angular momentum quantum number, and $R(r)$ and $S(\theta)$ are the radial and angular actions respectively. Substituting this ansatz into the simplified wave equation yields:
\begin{multline}
	g^{tt}(-iE)^2 \Phi + \frac{1}{\sqrt{-g}}\frac{\partial}{\partial r}\left(\sqrt{-g}g^{rr} i R'(r) \Phi \right) + g^{\phi \phi}(im)^2 \Phi \\
	+ \frac{1}{\sqrt{-g}}\frac{\partial}{\partial \theta}\left(\sqrt{-g}g^{\theta\theta} i S'(\theta) \Phi \right) + 2g^{t \phi}(-iE)(im) \Phi = 0
\end{multline}
Carrying out the derivatives on the product terms:
\begin{multline}
	-g^{tt}E^2 \Phi + \frac{i\Phi}{\sqrt{-g}}\frac{\partial}{\partial r}(\sqrt{-g}g^{rr})R'(r) + g^{rr}\left[i R''(r) - (R'(r))^2\right]\Phi - m^2 g^{\phi \phi}\Phi \\
	+ \frac{i\Phi}{\sqrt{-g}}\frac{\partial}{\partial \theta}(\sqrt{-g}g^{\theta\theta})S'(\theta) + g^{\theta\theta}\left[i S''(\theta) - (S'(\theta))^2\right]\Phi + 2mE g^{t \phi}\Phi = 0
\end{multline}
In the WKB approximation, the amplitude is assumed to vary slowly compared to the phase. We consider the leading order terms (real part), neglecting the second derivatives $R''(r)$ and $S''(\theta)$ compared to the squared first derivatives $(R')^2$ and $(S')^2$. Dividing by $\Phi$, we isolate the real part:
\begin{equation}
	-g^{tt}E^2 - g^{rr}\left(\frac{\partial R(r)}{\partial r} \right)^2 - m^2 g^{\phi \phi} - g^{\theta\theta}\left(\frac{\partial S(\theta)}{\partial \theta} \right)^2 + 2mE g^{t\phi} = 0
\end{equation}
We can now solve for the radial momentum $k_r \equiv \frac{\partial R}{\partial r}$. Rearranging the terms gives:
\begin{equation}
	g^{rr}\left(\frac{\partial R}{\partial r}\right)^2 = -g^{tt}E^2 - m^2g^{\phi \phi} - g^{\theta\theta}\left(\frac{\partial S}{\partial \theta} \right)^2 + 2mE g^{t\phi}
\end{equation}
Defining the generalized angular momentum $k_\theta \equiv \frac{\partial S}{\partial \theta}$, we arrive at the final expression for the radial wavenumber squared:
\begin{equation}
	k_{r}^2 = \frac{1}{g^{rr}}\left[ -g^{tt}E^2 - m^2g^{\phi \phi} - g^{\theta\theta}k_{\theta}^2 + 2mE g^{t\phi} \right]
\end{equation}
This expression serves as the starting point for calculating the density of states in the brick wall formalism.

\section{Free Energy and Entropy}\label{Free energy}

In the semiclassical approximation, the free energy of a scalar field in a stationary axisymmetric background is determined by the canonical ensemble partition function. The total free energy $\beta F = -\ln Z$ for bosons is given by the summation over all quantum numbers:
\begin{equation}
	\beta F = \sum_{E,m} \ln\left[1 - e^{-\beta(E - m\Omega_{H})}\right]
\end{equation}
Replacing the summation with an integration over the continuous energy spectrum and the density of states $\frac{d\Gamma(E)}{dE}$, we write:
\begin{align}
	\beta F &= \int dm \int_{0}^{\infty} d\Gamma(E) \ln\left[1 - e^{-\beta(E - m\Omega_{H})}\right] \nonumber \\
	&= \int dm \left( \left[\Gamma(E) \ln(1 - e^{-\beta(E - m\Omega_{H})})\right]_{0}^{\infty} - \int_{0}^{\infty} \Gamma(E) \frac{\beta e^{-\beta(E - m\Omega_{H})}}{1 - e^{-\beta(E - m\Omega_{H})}} dE \right) \nonumber \\
	&= -\beta \int dm \int_{0}^{\infty} dE \frac{\Gamma(E)}{e^{\beta(E - m\Omega_{H})} - 1}
\end{align}
where the boundary term vanishes. To proceed, we require the expression for the total number of modes $\Gamma(E,m)$ with energy less than $E$ for a fixed azimuthal quantum number $m$. In the WKB approximation, this is obtained by integrating the volume of the phase space within the shell $r_{H} + \epsilon \le r \le L$:
\begin{equation}
	\Gamma(E,m) = \frac{1}{\pi} \int d\theta d\phi \int_{r_{H} + \epsilon}^{L} dr \int dk_{\theta} k_{r}(r, \theta, E, m)
\end{equation}
Substituting the expression for the radial momentum $k_r$ derived from the Hamilton-Jacobi equation:
\begin{equation}
	\Gamma(E,m) = \frac{1}{\pi} \int d\theta d\phi \int_{r_{H} + \epsilon}^{L} \frac{dr}{\sqrt{g^{rr}}} \int dk_{\theta} \sqrt{-g^{tt}E^2 - m^2g^{\phi \phi} - g^{\theta\theta}k_{\theta}^2 + 2mE g^{t\phi}}
\end{equation}
The integration over $k_{\theta}$ is performed over the classically allowed region where $k_r^2 \ge 0$. Let us define the effective potential term $A^2$:
\begin{equation}
	A^2 \equiv -g^{tt}E^2 - m^2g^{\phi \phi} + 2mE g^{t\phi}
\end{equation}
The condition $k_r^2 \ge 0$ implies $g^{\theta\theta}k_\theta^2 \le A^2$. The integration over $k_\theta$ essentially yields the area of a semi-circle:
\begin{align}
	\int dk_{\theta} \sqrt{A^2 - g^{\theta\theta}k_{\theta}^2} &= \frac{1}{\sqrt{g^{\theta\theta}}} \int_{0}^{A} \sqrt{A^2 - x^2} \, dx \notag\\
	&= \frac{1}{\sqrt{g^{\theta\theta}}}\left|\frac{1}{2}\left(x\sqrt{A^2 - x^2} + A^2 \tan^{-1}\frac{x}{\sqrt{A^2-x^2}}\right)\right|_{0}^{A}= \frac{1}{\sqrt{g^{\theta\theta}}} \left(\frac{\pi A^2}{4}\right)
\end{align}
Substituting this back into the expression for $\Gamma(E,m)$, we obtain:
\begin{equation}
	\Gamma(E,m) = \frac{1}{4} \int d\theta d\phi \int_{r_{H} + \epsilon}^{L} dr \frac{-g^{tt}E^2 - m^2g^{\phi \phi} + 2mE g^{t\phi}}{\sqrt{g^{rr}g^{\theta\theta}}}
\end{equation}
Finally, substituting $\Gamma(E,m)$ into the free energy integral gives:
\begin{equation}
	F = -\frac{1}{4} \int dm \int d\theta d\phi \int_{r_{H} + \epsilon}^{L} dr \frac{1}{\sqrt{g^{rr}g^{\theta\theta}}} \int_{0}^{\infty} dE \frac{-g^{tt}E^2 - m^2g^{\phi \phi} + 2mE g^{t\phi}}{e^{\beta(E - m\Omega_{H})} - 1}
\end{equation}
This integral exhibits a pole at the superradiance limit $E = m\Omega_{H}$, where the denominator vanishes. To handle this singularity and isolate the superradiant contributions, we split the energy integration into two regimes: the superradiant interval $0 \le E \le m\Omega_{H}$ and the normal interval $m\Omega_{H} < E < \infty$:
\begin{multline}
	F = \underbrace{-\frac{1}{4} \int \dots \int_{0}^{m\Omega_{H}} dE \frac{-g^{tt}E^2 - m^2g^{\phi \phi} + 2mE g^{t\phi}}{e^{\beta(E - m\Omega_{H})} - 1}}_{F_{\text{SR}}} \\
	\underbrace{-\frac{1}{4} \int \dots \int_{m\Omega_{H}}^{\infty} dE \frac{-g^{tt}E^2 - m^2g^{\phi \phi} + 2mE g^{t\phi}}{e^{\beta(E - m\Omega_{H})} - 1}}_{F_{\text{NSR}}}
\end{multline}
where $F_{\text{SR}}$ and $F_{\text{NSR}}$ denote the free energy contributions from superradiant and non-superradiant modes, respectively. We now evaluate the contribution from the non-superradiant modes ($E > m\Omega_{H}$). Starting from the general expression for free energy, we perform a change of variables $E' = E - m\Omega_{H}$ to shift the lower integration limit to zero.
\begin{align}
	F_{\text{NSR}} &= -\frac{1}{4}\int dm \int d\theta d \phi \int_{r_{H} + \epsilon}^{L} \frac{dr}{\sqrt{g^{rr}g^{\theta\theta}}} \int_{m\Omega_{H}}^{\infty} dE\frac{\left(-g^{tt}E^{2} -m^{2}g^{\phi \phi} + 2mEg^{t\phi}\right)}{e^{\beta(E - m\Omega_{H})} - 1} \nonumber \\
	&= -\frac{1}{4}\int dm \int d\theta d \phi \int_{r_{H} + \epsilon}^{L} \frac{dr}{\sqrt{g^{rr}g^{\theta\theta}}} \int_{0}^{\infty} dE'\frac{-g^{tt}(E' + m\Omega_{H})^2 -m^{2}g^{\phi \phi} + 2m(E' + m\Omega_{H})g^{t\phi}}{e^{\beta E'} - 1}
\end{align}
To facilitate integration, we swap the order of integration between $m$ and $E'$. The limits for the $m$ integration are determined by the condition that the radial momentum $k_r$ must be real, which implies that the term inside the square root of the density of states must be non-negative. This defines the roots $b_{\pm}$ of the quadratic equation in $m$.
\begin{align}
	F_{\text{NSR}} &= -\frac{1}{4}\int d\theta d \phi \int_{r_{H} + \epsilon}^{L} \frac{dr}{\sqrt{g^{rr}g^{\theta\theta}}} \int_{0}^{\infty} \frac{dE'}{e^{\beta E'} - 1} \nonumber \\
	&\quad \times \int_{b_{+}}^{b_{-}} dm \left[ -g^{tt}(E' + m\Omega_{H})^2 -m^{2}g^{\phi \phi} + 2m(E' + m\Omega_{H})g^{t\phi}\right]
\end{align}
where $b_{\pm}$ are the roots of the quadratic equation:
\[
g^{tt}(E' + m\Omega_{H})^2 + m^2g^{\phi \phi} - 2m(E' + m\Omega_{H})g^{t\phi} = 0
\]
Absorbing the negative sign by reversing the limits of the $m$-integration ($b_{-} \to b_{+}$), and grouping the terms by powers of $m$, we obtain:
\begin{align}
	F_{\text{NSR}} &= \frac{1}{4}\int d\theta d \phi \int_{r_{H} + \epsilon}^{L} \frac{dr}{\sqrt{g^{rr}g^{\theta\theta}}} \int_{0}^{\infty} \frac{dE'}{e^{\beta E'} - 1} \nonumber \\
	&\quad \times \int_{b_{-}}^{b_{+}} dm \left[ (g^{tt}\Omega_{H}^{2} + g^{\phi \phi} - 2\Omega_{H} g^{t\phi})m^2 + 2E'(g^{tt}\Omega_{H}-g^{t\phi})m + g^{tt}{E'}^{2}\right]
\end{align}
Using the relation $\mathcal{D} = g_{tt}g_{\phi\phi} - g_{t\phi}^2$, $\Omega_{H}=-\frac{g_{t\phi}}{g_{\phi\phi}}$ and the inverse metric identities \eqref{inv-metric}, the coefficients simplify significantly. The quadratic integrand reduces to:
\begin{equation}
	F_{\text{NSR}} = \frac{1}{4}\int d\theta d \phi \int_{r_{H} + \epsilon}^{L} \frac{dr}{\sqrt{g^{rr}g^{\theta\theta}}} \int_{0}^{\infty} \frac{dE'}{e^{\beta E'} - 1}\int_{b_{-}}^{b_{+}} dm \left[ \frac{1}{g_{\phi\phi}}m^2 + \frac{g_{\phi\phi}}{\mathcal{D}}{E'}^{2} \right]
\end{equation}
Integrating this result with respect to $m$ and substituting the limit we arrive at:
\begin{align}
F_{NSR} &= \frac{1}{4}\int d\theta d \phi \int_{r_{H} + \epsilon}^{L} dr \frac{1}{\sqrt{g^{rr}g^{\theta\theta}}}  \int_{0}^{\infty} d{E'}\frac{1}{e^{\beta E'} - 1}\left[-\frac{4}{3}\frac{{E'}^3}{\left\{\mathcal{(-D)}\right\}^{3/2}}(g_{\phi\phi})^2\right]\notag\\
&= -\frac{1}{3}\int d\theta d \phi \int_{r_{H} + \epsilon}^{L} dr \frac{1}{\sqrt{g^{rr} g^{\theta\theta}}} \frac{(g_{\phi\phi})^2}{\left\{\mathcal{(-D)}\right\}^{3/2}}\ \int_{0}^{\infty} d{E'}\frac{{E'}^{3} }{e^{\beta E'} - 1}\notag\\
&= -\frac{\Gamma(4)\zeta(4)}{3\beta^4} \int d\theta d \phi \int_{r_{H} + \epsilon}^{L} dr \frac{1}{\sqrt{g^{rr} g^{\theta\theta}}} \frac{(g_{\phi\phi})^2}{\mathcal{(-D)}^{3/2}}\notag\\
&= -\frac{\Gamma(4)\zeta(4)}{3\beta^4} \int d\theta d \phi \int_{r_{H} + \epsilon}^{L} dr \frac{\left(g_{_{rr}} g_{_{\theta\theta}}\right)^{1/2}}{\mathcal{(-D)}^{3/2}}(g_{\phi\phi})^2\label{FNSR}
\end{align}
To ensure the reality of the radial wave number $k_r$, we imposed the condition $k_r^2 \ge 0$. This requires the term under the square root in the density of states to be non-negative:
\begin{equation}
	-g^{tt}E^2 - m^2 g^{\phi \phi} - g^{\theta\theta}k_{\theta}^2 + 2mE g^{t\phi} \ge 0
\end{equation}
Since $g^{\theta\theta}k_{\theta}^2 \ge 0$, a necessary condition for the existence of any allowed mode is that the maximum value of the effective potential must be non-negative:
\begin{equation}
	-g^{tt}E^2 - m^2 g^{\phi \phi} + 2mE g^{t\phi} \ge 0
\end{equation}
Multiplying by $-1$ reverses the inequality:
\begin{equation}
	g^{tt}E^2 + m^2 g^{\phi \phi} - 2mE g^{t\phi} \le 0
\end{equation}
Substituting the shifted energy $E = E' + m\Omega_H$ (where $E' > 0$ for NSR modes) into this inequality:
\begin{equation}
	g^{tt}(E' + m\Omega_{H})^2 + m^2 g^{\phi \phi} - 2m(E' + m\Omega_{H})g^{t\phi} \le 0
\end{equation}
Expanding and collecting terms in powers of $m$:
\begin{equation}
	\left(g^{tt}\Omega_{H}^{2} + g^{\phi \phi} - 2\Omega_{H} g^{t\phi}\right)m^2 + 2E'(g^{tt}\Omega_{H} - g^{t\phi})m + g^{tt}{E'}^{2} \le 0
\end{equation}
We simplify the coefficients using the definition $\Omega_H = -g_{t\phi}/g_{\phi\phi}$ and the determinant relation $\mathcal{D} = g_{tt}g_{\phi\phi} - g_{t\phi}^2$.
For the $m^2$ coefficient, algebraic simplification yields:
\begin{equation}
	g^{tt}\Omega_{H}^{2} + g^{\phi \phi} - 2\Omega_{H} g^{t\phi} = \frac{1}{\mathcal{D}g_{\phi\phi}}\left( g_{\phi\phi}\Omega_H^2 g_{\phi\phi} + g_{tt}g_{\phi\phi} + 2\Omega_H g_{t\phi}g_{\phi\phi} \right)=\frac{\mathcal{D}}{\mathcal{D}g_{\phi\phi}}
\end{equation}
Crucially, the coefficient for the linear term in $m$ vanishes identically upon substitution of the inverse metric components:
\begin{equation}
	2E'(g^{tt}\Omega_{H} - g^{t\phi}) = 2E'\left( \frac{g_{\phi\phi}}{\mathcal{D}}\left(-\frac{g_{t\phi}}{g_{\phi\phi}}\right) - \left(-\frac{g_{t\phi}}{\mathcal{D}}\right) \right) = 0
\end{equation}
Thus, the inequality simplifies drastically to:
\begin{equation}
	\frac{1}{g_{\phi\phi}}m^2 + \frac{g_{\phi\phi}}{\mathcal{D}}{E'}^{2} \le 0
\end{equation}
Solving for $m$, and recalling that $\mathcal{D} < 0$ (so $-\mathcal{D} > 0$) and $g_{\phi\phi} > 0$ outside the horizon:
\begin{equation}
	m^2 \le - \frac{(g_{\phi\phi})^2}{\mathcal{D}} {E'}^2 = \frac{(g_{\phi\phi})^2}{-\mathcal{D}} {E'}^2
\end{equation}
This yields the symmetric integration limits $b_{\pm}$:
\begin{equation}
	-\sqrt{\frac{{E'}^{2}}{-\mathcal{D}}}g_{\phi\phi} \le m \le \sqrt{\frac{{E'}^{2}}{-\mathcal{D}}}g_{\phi\phi}
\end{equation}
These correspond precisely to the limits $b_{-}$ and $b_{+}$ utilized in the evaluation of $F_{\text{NSR}}$. We now turn our attention to the superradiant modes, defined by the energy range $0 \le E \le m\Omega_{H}$. To facilitate the evaluation, we introduce the dimensionless variable $x$ such that $E = m\Omega_{H}x$, where $0 \le x \le 1$. Substituting this into the expression for $F_{\text{SR}}$:
\begin{align*}
F_{SR} &= -\frac{1}{4}\int dm \int d\theta d \phi\int_{r_{H} + \epsilon}^{L} dr \frac{1}{\sqrt{g^{rr}g^{\theta\theta}}} \int_{0}^{m\Omega_{H}} dE\frac{\left(-g^{tt}E^2 -m^2g^{\phi \phi} + 2mEg^{t\phi}\right)}{e^{\beta(E - m\Omega_{H})} - 1}\\
\\
&= -\frac{1}{4}\int dm \int d\theta d \phi \int_{r_{H} + \epsilon}^{L} dr \frac{1}{\sqrt{g^{rr}g^{\theta\theta}}} \int_{0}^{1} m\Omega_{H} dx\frac{-g^{tt}(m\Omega_{H}x)^2 -m^2 g^{\phi \phi} + 2m(m\Omega_{H}x)g^{t\phi}}{e^{\beta m\Omega_{H}(x - 1)} - 1}\\
\\
&= -\frac{\Omega_{H}}{4}\int dm \int d\theta d \phi \int_{r_{H} + \epsilon}^{L} dr \frac{1}{\sqrt{g^{rr}g^{\theta\theta}}} \int_{0}^{1} m^3 dx\frac{-g^{tt}(\Omega_{H}x)^2 - g^{\phi \phi} + 2(\Omega_{H}x)g^{t\phi}}{e^{\beta m\Omega_{H}(x - 1)} - 1}\\
\\
&= -\frac{\Omega_{H}}{4} \int_{0}^{1}dx \int d\theta d \phi \int_{r_{H} + \epsilon}^{L} dr \frac{1}{\sqrt{g^{rr}g^{\theta\theta}}} \int dm\frac{\Omega^2m^3}{e^{\beta m\Omega_{H}(x - 1)} - 1}
\end{align*}
In the last step we used $\Omega^2 = -g^{tt}(\Omega_{H}x)^2 - g^{\phi \phi} + 2(\Omega_{H}x)g^{t\phi}$. The limit for $m$-integral can be determined as follows:
\begin{align}
-g^{tt}(m\Omega_{H}x)^2 -m^2g^{\phi \phi} + 2m(m\Omega_{H}x)g^{t\phi} &\ge 0\notag\\
m^{2}\{-g^{tt}(\Omega_{H}x)^2 - g^{\phi \phi} + 2(\Omega_{H}x)g^{t\phi}\} &\ge 0 \tag{this suggests $m^2\ge 0$}\\
(g^{tt}\Omega_{H}^{2})x^{2} - (2\Omega_{H}g^{t\phi})x + g^{\phi \phi}&\le 0\notag\\
\frac{1}{\mathcal{D}}\left[(g_{\phi\phi}\Omega_{H}^{2})x^{2} + (2\Omega_{H}g_{t\phi})x + g_{tt}\right]&\le 0\label{x-lim}
\end{align}

Although the superradiant regime is defined by frequencies $0 \le E \le m\Omega_H$ (or $0 \le x \le 1$), the scalar field modes must also satisfy the kinematic constraint imposed by the background geometry. Given that $m$ is real valued for $\mathcal{D}<0$, the allowed values of $x$ is restricted and there is an upper bound on the integration variable $x$. We first note that $\left[(g_{\phi\phi}\Omega_{H}^{2})x^{2}+ (2\Omega_{H}g_{t\phi})x + g_{tt}\right]\ge 0$ is only satisfied in the region $r_{H}<r<r_{\rm erg}$ for $0<x<1$. Hence the bound on $x$ in this region can be estimated as following:
\begin{align*}
	 0\le x \le &\frac{-2g_{t\phi}\Omega_{H} - \sqrt{(2g_{t\phi}\Omega_{H})^{2} - 4g_{tt} g_{\phi \phi }\Omega_{H}^{2}}}{2g_{\phi\phi}\Omega_{H}^{2}}\\
0\le x &\le -\frac{(g_{t\phi} + \sqrt{\mathcal{-D}})}{g_{\phi\phi}\Omega_{H}} \le 1\tag{using $\mathcal{-D}=  (g_{t\phi})^2-g_{tt} g_{\phi \phi }$}\\
0\le x &\le 1 + \frac{\sqrt{\mathcal{-D}}}{g_{t\phi}}
\tag{using $\Omega_{H} = -\frac{g_{t\phi}}{g_{\phi\phi}}$}
\end{align*}
Outside the horizon but inside the ergosphere, $g_{t\phi} < 0,\Delta>0$, which implies that the term $\frac{\sqrt{-\mathcal{D}}}{g_{t\phi}}$ is negative. Thus, the kinematic upper limit $\alpha$ is slightly less than 1. Substituting this back into the free energy expression:
$$F_{SR} = -\frac{\Omega_{H}}{4} \int_{0}^{\alpha}dx \int d\theta d \phi \int_{r_{H} + \epsilon}^{L} dr \frac{\Omega^2}{\sqrt{g^{rr}g^{\theta\theta}}} \int_{0}^{\infty} dm\frac{m^3}{e^{\beta m\Omega_{H}(x - 1)} - 1}$$
This integral is formally divergent because the exponent coefficient $\beta \Omega_H (x-1)$ is negative for $x < 1$. A quick check can be performed (e.g., using Mathematica) to verify this behavior:
\begin{verbatim}
	Integrate[m^3/(Exp[-k m] - 1), {m, 0, Infinity}, Assumptions -> k > 0]
\end{verbatim}
The output confirms that the integral diverges. Therefore, following Ref.~\cite{ho1997entropy}, we introduce an exponential regulator $e^{-\epsilon_m m}$ and take the limit $\epsilon_{m} \to 0$ at the end of the calculation:
\begin{align*}
F_{SR} &= -\frac{\Omega_{H}}{4} \int_{0}^{\alpha}dx \int d\theta d \phi \int_{r_{H} + \epsilon}^{L} dr \frac{\Omega^2}{\sqrt{g^{rr}g^{\theta\theta}}} \int_{0}^{\infty} dm\frac{m^3}{e^{\beta m\Omega_{H}(x - 1)} - 1}e^{-\epsilon_m m}\\
&= \frac{\Omega_{H}}{4} \int_{0}^{\alpha}dx \int d\theta d \phi \int_{r_{H} + \epsilon}^{L} dr \frac{\Omega^2}{\sqrt{g^{rr}g^{\theta\theta}}} \left[ \frac{\Gamma(4)\zeta(4)}{\{\beta \Omega_{H}(x - 1)\}^4}+\frac{\Gamma(4)}{\epsilon_m^4}\right]
\intertext{The integral with the regulator can be expressed in terms of the PolyGamma function. To extract the leading order coefficient, we series expanded the polygamma function $\phi^{(3)}\left(\frac{\epsilon_m}{\beta\Omega_{H}(1-x)}\right)$ near $\epsilon_m=0$ and kept terms up to first order:}
&\approx\frac{\Gamma(4)\zeta(4)}{4\beta^{4}\Omega_{H}^{3}} \int d\theta d \phi \int_{r_{H} + \epsilon}^{L} dr\frac{1}{\sqrt{g^{rr}g^{\theta\theta}}}\int_{0}^{\alpha}dx\frac{\Omega^2}{(x - 1)^4}\\
\intertext{using $\Omega^2 = -g^{tt}(\Omega_{H}x)^2 - g^{\phi \phi} + 2(\Omega_{H}x)g^{t\phi}$}
&= -\frac{\Gamma(4)\zeta(4)}{4\beta^{4}\Omega_{H}^{3}} \int d\theta d \phi \int_{r_{H} + \epsilon}^{L} dr\frac{1}{\sqrt{g^{rr}g^{\theta\theta}}}\int_{0}^{\alpha}dx\frac{g^{tt}\Omega_{H}^2x^2 - 2\Omega_{H}g^{t\phi}x + g^{\phi \phi} }{(x - 1)^4}\\
&= -\frac{\Gamma(4)\zeta(4)}{4\beta^{4}\Omega_{H}^{3}} \int d\theta d \phi \int_{r_{H} + \epsilon}^{L} dr\sqrt{g_{rr}g_{\theta\theta}}\int_{0}^{\alpha}dx\frac{g_{\phi \phi}\Omega_{H}^2x^2 + 2\Omega_{H}g_{t\phi}x + g_{tt} }{\mathcal{D}(x - 1)^4}\\
&= -\frac{\Gamma(4)\zeta(4)}{4\beta^{4}\Omega_{H}^{3}} \int d\theta d \phi \int_{r_{H} + \epsilon}^{L} dr\sqrt{g_{rr}g_{\theta\theta}}\int_{0}^{\alpha}dx~g_{\phi \phi}\Omega_{H}^2\frac{x^2}{\mathcal{D}(x - 1)^4} + 2\Omega_{H}g_{t\phi}\frac{x}{\mathcal{D}(x - 1)^4} + g_{tt}\frac{1 }{\mathcal{D}(x - 1)^4}\\
&= -\frac{\Gamma(4)\zeta(4)}{4\beta^{4}\Omega_{H}^{3}} \int d\theta d \phi \int_{r_{H} + \epsilon}^{L} dr\frac{\sqrt{g_{rr}g_{\theta\theta}}}{\mathcal{D}}\left[~g_{\phi \phi}\Omega_{H}^2\left|\frac{-3 x^2+3 x-1}{3 (x-1)^3}\right|_{0}^{\alpha} + 2\Omega_{H}g_{t\phi}\left|\frac{1-3 x}{6 (x-1)^3}\right|_{0}^{\alpha} + g_{tt}\left|-\frac{1}{3 (x-1)^3}\right|_{0}^{\alpha}\right]\\
\end{align*}
The $F_{SR}$ is composed of several small pieces, therefore let us simplify each one of them to extract the leading order contribution:
\begin{align*}
I &= g_{\phi \phi}\Omega_{H}^2\frac{-3 \alpha^2+3 \alpha-1}{3 (\alpha-1)^3} + 2\Omega_{H}g_{t\phi}\frac{1-3\alpha}{6 (\alpha-1)^3} - g_{tt}\frac{1}{3 (\alpha-1)^3} - \frac{g_{\phi \phi}\Omega_{H}^2}{3} +\frac{\Omega_{H}g_{t\phi}}{3} - \frac{g_{tt}}{3}\\
&= g_{\phi \phi}\Omega_{H}^2\frac{-3 \alpha^2+3 \alpha-1}{3 (\alpha-1)^3} + \Omega_{H}g_{t\phi}\frac{1-3\alpha}{3 (\alpha-1)^3} - g_{tt}\frac{1}{3 (\alpha-1)^3} + \frac{\Omega_{H}g_{t\phi}-g_{\phi \phi}\Omega_{H}^2-g_{tt}}{3}\\
&= g_{\phi \phi}\Omega_{H}^2\frac{-3( \alpha^2 - \alpha)-1}{3 (\alpha-1)^3} + \Omega_{H}g_{t\phi}\frac{(3-2)-3\alpha}{3 (\alpha-1)^3} - g_{tt}\frac{1}{3 (\alpha-1)^3} + \frac{\Omega_{H}g_{t\phi}-g_{\phi \phi}\Omega_{H}^2-g_{tt}}{3}\\
&= g_{\phi \phi}\Omega_{H}^2\frac{-3( \alpha-1)^2 - 3(\alpha-1) -1}{3 (\alpha-1)^3} + \Omega_{H}g_{t\phi}\frac{-3(\alpha-1) -2}{3 (\alpha-1)^3} - g_{tt}\frac{1}{3 (\alpha-1)^3} + \frac{\Omega_{H}g_{t\phi}-g_{\phi \phi}\Omega_{H}^2-g_{tt}}{3}\\
&= -3\frac{g_{\phi \phi}\Omega_{H}^2}{3 (\alpha-1)} -3 \frac{\Omega_{H}g_{t\phi} +g_{\phi \phi}\Omega_{H}^2 }{3 (\alpha-1)^2} - \frac{ g_{\phi \phi}\Omega_{H}^2 + 2\Omega_{H}g_{t\phi} + g_{tt}}{3 (\alpha-1)^3} + \frac{\Omega_{H}g_{t\phi}-g_{\phi \phi}\Omega_{H}^2-g_{tt}}{3}\\
&= I_{1} + I_{2} + I_{3} + I_{4}
\end{align*}
These need to be simplified using the expression of $\alpha$ that we found earlier:
\begin{align*}
\alpha &= -\frac{g_{t\phi} + \sqrt{(g_{t\phi})^2 - g_{tt} g_{\phi \phi }}}{g_{\phi\phi}\Omega_{H}} = \frac{g_{t\phi} + \sqrt{\mathcal{-D}}}{g_{t\phi}} \implies \alpha -1 = \frac{\sqrt{\mathcal{-D}}}{g_{t\phi}}
\end{align*}
The term $I_{1}/\Omega_{H}^{3}$ can be simplified to:
\begin{align*}
\frac{I_{1}}{\Omega_{H}^3} &= -3\frac{g_{\phi \phi}\Omega_{H}^2}{3 \Omega_{H}^3(\alpha-1)}\\
&= -3\frac{g_{\phi \phi}}{3\Omega_{H}(\alpha-1)}\\
&= \frac{3(g_{\phi \phi})^{2}}{3\sqrt{\mathcal{-D}}} = -3\frac{\mathcal{D}}{3(\mathcal{-D})^{3/2}}(g_{\phi \phi})^{2} = -3\frac{I_{3}}{\Omega_{H}^{3}}
\tag{using $\Omega_{H} = -\frac{g_{t\phi}}{g_{\phi\phi}}$}
\end{align*}
The second term $I_{2}/\Omega_{H}^{3}$ vanishes:
\begin{align*}
\frac{I_{2}}{\Omega_{H}^3} &= -3 \frac{\Omega_{H}g_{t\phi} +g_{\phi \phi}\Omega_{H}^2 }{3\Omega_{H}^3 (\alpha-1)^2}\\
&=\frac{g_{t\phi} +g_{\phi \phi}\Omega_{H}}{ \Omega_{H}^2(\alpha-1)^2}\\
&=\frac{g_{t\phi} -g_{t\phi}}{ \Omega_{H}^2(\alpha-1)^2}= 0
\end{align*}
The third term $I_{3}/\Omega_{H}^{3}$ becomes:
\begin{align*}
\frac{I_{3}}{\Omega_{H}^{3}} &=- \frac{ g_{\phi \phi}\Omega_{H}^2 + 2\Omega_{H}g_{t\phi} + g_{tt}}{3 \Omega_{H}^{3}(\alpha-1)^3}\\
&=-\frac{(g_{t\phi})^2 -2(g_{t\phi})^2 + g_{tt}g_{\phi\phi} }{3 \Omega_{H}^{3}g_{\phi\phi}(\sqrt{\mathcal{-D}})^3}(g_{t\phi})^{3}
\tag{using $\Omega_{H} = -\frac{g_{t\phi}}{g_{\phi\phi}}$}\\
&=\frac{\mathcal{D}}{3(\mathcal{-D})^{3/2}}(g_{\phi\phi})^{2}
\tag{$\mathcal{D} = -(g_{t\phi})^2 + g_{tt} g_{\phi \phi } $}
\end{align*}
while the last term $I_{4}/\Omega_{H}^{3}$ becomes:
\begin{align*}
\frac{I_{4}}{\Omega_{H}^3}&=\frac{\Omega_{H}g_{t\phi}-g_{\phi \phi}\Omega_{H}^2-g_{tt}}{3\Omega_{H}^3}\\
&=\frac{\Omega_{H}g_{t\phi}-g_{\phi \phi}\Omega_{H}^2-g_{tt}}{3\Omega_{H}^3}\\
&=\frac{-(g_{t\phi})^{2}-(g_{t\phi})^{2}-g_{tt}g_{\phi\phi}}{3g_{\phi\phi}\Omega_{H}^3}\\
&=\frac{2(g_{t\phi})^{2}+g_{tt}g_{\phi\phi}}{3(g_{t\phi})^3}(g_{\phi\phi})^{2}\\
&= \frac{3(g_{t\phi})^{2}+\mathcal{D}}{3(g_{t\phi})^3}(g_{\phi\phi})^{2}
= \frac{(g_{\phi\phi})^{2}}{(g_{t\phi})} + \frac{\mathcal{D}}{3(g_{t\phi})^3}(g_{\phi\phi})^{2}
\end{align*}
Keeping only, $I_{1}$,$I_{2}$ \& $I_{3}$, the $F_{SR}$ can be given as:
\begin{align}
F_{SR}&\approx \frac{\Gamma(4)\zeta(4)}{6\beta^{4}} \int d\theta d \phi \int_{r_{H} + \epsilon}^{L} dr\frac{\sqrt{g_{rr}g_{\theta\theta}}}{\mathcal{D}}\frac{\mathcal{D}}{(\mathcal{-D})^{3/2}}(g_{\phi\phi})^{2}\notag\\
&=\frac{\Gamma(4)\zeta(4)}{6\beta^{4}} \int d\theta d \phi \int_{r_{H} + \epsilon}^{L} dr\frac{\sqrt{g_{rr}g_{\theta\theta}}}{(\mathcal{-D})^{3/2}}(g_{\phi\phi})^{2}\label{FSR}
\end{align}
We find that:
\begin{align*}
F_{SR} &\approx -\frac{1}{2}F_{NSR}
\end{align*}

\section{Applications to Specific Black Hole Geometries}

Having established a unified framework for calculating the free energy of scalar fields in stationary axisymmetric spacetimes, we now proceed to apply this formalism to specific black hole solutions. This serves to demonstrate the universality of our generalized master equations, allowing us to extract the entropy simply by substituting the relevant metric components ($g_{tt}, g_{\phi\phi}, g_{rr}, \dots$) and the determinant $\mathcal{D}$.

In the following subsections, we apply our results to two distinct cases of increasing complexity. First, we consider the Kerr-Newman-AdS spacetime to verify the consistency of our method in the presence of a cosmological constant. Subsequently, we extend the analysis to a more non-trivial background: a Kerr-Newman-AdS black hole surrounded by quintessence and a cloud of strings, investigating how these exotic matter fields influence the thermodynamic area law.
\subsection{Kerr Newman}
the Kerr-Newman black hole, which we will use to validate our results, the metric components in Boyer-Lindquist coordinates are given by:\cite{newman1965metric}
\begin{equation}
	ds^2 = - \left( \frac{\Delta - a^2\sin^2\theta}{\rho^2} \right) dt^2 + \rho^2d\theta^2 + \frac{\rho^2}{\Delta}{dr^2} + \frac{\Sigma\sin^2\theta}{\rho^2}d\phi^2 - \frac{2a\sin^2\theta (2GMr-Q^2)}{\rho^2}dt d\phi\label{kerr-Newman-metric}
\end{equation}
where the parameters are defined as:
\begin{align}
	\Delta &= a^2 + r^2 - 2GMr + Q^2 \nonumber \\
	\Sigma &= (r^2 + a^2)^2 - a^2 \Delta \sin^2\theta \nonumber \\
	\rho^2 &=  r^2 + a^2\cos^2\theta \nonumber \\
	\sqrt{-g} &= \rho^2 \sin\theta
\end{align}
The inverse metric corresponding to above can be given as following:
\begin{equation}
g^{\mu\nu}_{\text{KN}} = 
\begin{pmatrix}
	-\frac{\Sigma}{\rho^2\Delta} & 0 & 0 & -\frac{a(2GMr-Q^2)}{\rho^2\Delta} \\ 
	0 & \frac{\Delta}{\rho^2} & 0 & 0 \\ 
	0 & 0 & \frac{1}{\rho^2} & 0 \\ 
	-\frac{a(2GMr-Q^2)}{\rho^2\Delta} & 0 & 0 &  \frac{\Delta - a^2\sin^2\theta}{\rho^2\Delta\sin^2\theta}
\end{pmatrix}
\end{equation}
The generalized expression for $F_{NSR}$ in \eqref{FNSR}, and using $\mathcal{D}= \Delta \sin^2\theta$ and components $g_{rr}, g_{\theta\theta}, g_{\phi\phi}$ defined in Eq.~\eqref{kerr-Newman-metric}, we can evaluate it for this specific background. As $r \to r_H$, the function $\Delta(r)$ approaches zero, leading to coordinate singularity. We expand the integrand near the horizon $r_H$ to isolate the leading order behavior. Recall that $\Delta = (r - r_+)(r - r_-)$, so near the outer horizon $r_+ = r_H$:
$$\frac{1}{(r - r_{H})^2}F(r,\theta)= \frac{F(r_{_{H}},\theta)}{(r - r_{H})^2} +\frac{F'(r_{_{H}},\theta)}{(r - r_{H})} +\mathcal{O}(r - r_{H})$$
where $$F(r,\theta)= \frac{\left(g_{_{rr}} g_{_{\theta\theta}}\right)^{1/2}}{\left\{\mathcal{(-D)}\right\}^{3/2}}(g_{\phi\phi})^2(r-r_{_{H}})^{2}$$
using taylor expansion, the integrand approximates to:
\begin{align*}
	\frac{1}{(r-r_{_{H}})^{2}}\left[\frac{\left(g_{_{rr}} g_{_{\theta\theta}}\right)^{1/2}(g_{\phi\phi})^{2}}{\mathcal{(-D)}^{3/2}}(r-r_{_{H}})^{2} \right] &\approx \left|\frac{\sqrt{\frac{\rho^4}{\Delta}}\frac{\sin^4\theta(r^2 + a^2)^{4}}{\rho^{4}}}{\left\{\Delta\sin^2\theta\right\}^{3/2}}(r-r_{H})^{2} \right|_{r=r_{H}}\frac{1}{(r-r_{H})^2}\\
	&= \frac{\sin^{4}\theta (r_{H}^2 + a^2)^{4}}{\rho_{H}^{2}\left\{(r_{H}-r_{-})^{2}\sin^{3}\theta\right\}} \frac{1}{(r-r_{H})^2}\\
	&= \frac{(r_{H}^2 + a^2)^{4}\sin\theta}{(r_{H}^2 + a^2\cos^{2}\theta)\left\{(r_{H}-r_{-})\right\}^{2}}\frac{1}{(r-r_{H})^2}\\
	&= \frac{(r_{H}^2 + a^2)^{4}\sin\theta}{(r_{H}^2 + a^2\cos^{2}\theta)\left\{2(r_{H}-GM)\right\}^{2}}\frac{1}{(r-r_{H})^2}
\end{align*}

At the last step of approximation, we used:
\begin{align*}
	r_{+} - r_{-} &= \frac{-b + \sqrt{D}}{2a} - \frac{-b - \sqrt{D}}{2a}\\
	&= \frac{\sqrt{D}}{a} = \frac{2ar_{+} + b}{a} = 2r_{+} + \frac{b}{a}
\end{align*}
which leads to:
\begin{align*}
	\Delta=r^2-2GMr +(Q^2+a^2)=0\implies r_{+} - r_{-}=r_{H} -r_{-} &= 2r_{H} -2GMr 
\end{align*}
Using this approximation, the free energy integral near the horizon becomes:
\begin{align*}
	F_{NSR} &\approx -\frac{\Gamma(4)\zeta(4)}{3\beta^4} \int d\theta d \phi \int_{r_{H} + \epsilon}^{L} dr \frac{(r_{H}^2 + a^2)^{4}\sin\theta}{(r_{H}^2 + a^2\cos^{2}\theta)\left\{2(r_{H}-GM)\right\}^{2}}\frac{1}{(r-r_{H})^2}\\
	\\
	&=-\frac{\Gamma(4)\zeta(4)}{3\beta^4} \int d\theta d \phi \frac{(r_{H}^2 + a^2)^{4}\sin\theta}{(r_{H}^2 + a^2\cos^{2}\theta)\left\{2(r_{H}-GM)\right\}^{2}}\int_{r_{H} + \epsilon}^{L} dr \frac{1}{(r-r_{H})^2}\\
	\\
	&=\frac{\Gamma(4)\zeta(4)}{6\beta^4} \int d\theta d \phi \frac{(r_{H}^2 + a^2)^{4}\sin\theta}{(r_{H}^2 + a^2\cos^{2}\theta)\left\{(r_{H}-GM)\right\}^{2}}\left(\frac{1}{r-r_{H}}\right)_{r_{H} + \epsilon}^{L}\\
	\\
	&=-\frac{\zeta(4)}{\beta^4} \int d\theta d \phi \frac{(r_{H}^2 + a^2)^{4}\sin\theta}{(r_{H}^2 + a^2\cos^{2}\theta)\left\{(r_{H}-GM)\right\}^{2}}\left[\frac{\delta}{\epsilon(\epsilon + \delta)}\right] \tag{using $L = r_{H} + \epsilon + \delta$}
\end{align*}
We will use the generalized expression for $F_{SR}$ in \eqref{FSR}, and similarly, using the metric determinant $\mathcal{D}$ and components $g_{rr}, g_{\theta\theta}, g_{\phi\phi}$ defined in Eq.~\eqref{kerr-Newman-metric}, we can evaluate it as following:
\begin{align}
	F_{SR}&= \frac{\Gamma(4)\zeta(4)}{6\beta^{4}} \int d\theta d \phi \int_{r_{H} + \epsilon}^{L} dr\frac{\sqrt{g_{rr}g_{\theta\theta}}}{(\mathcal{-D})^{3/2}}(g_{\phi\phi})^{2}\notag \\
	&\approx -\frac{\Gamma(4)\zeta(4)}{6\beta^{4}} \int d\theta d \phi \frac{(r_{H}^2 + a^2)^{4}\sin\theta}{(r_{H}^2 + a^2\cos^{2}\theta)\left\{2(r_{H}-GM)\right\}^{2}}\left(\frac{1}{r-r_{H}}\right)_{r_{H} + \epsilon}^{L}\notag\\
	&=\frac{\zeta(4)}{2\beta^4} \int d\theta d \phi \frac{(r_{H}^2 + a^2)^{4}\sin\theta}{(r_{H}^2 + a^2\cos^{2}\theta)\left\{(r_{H}-GM)\right\}^{2}}\left[\frac{\delta}{\epsilon(\epsilon + \delta)}\right] \tag{using $L = r_{H} + \epsilon + \delta$ and $\Gamma(4)=6$}
\end{align}
Depending on the limit in which $r-$integral is carried out, the super-radiant contribution varies. This leads to modification in total free energy. The super-radiant and non-superradiant modes of free energy are now both well understood. If we carried out first with the thin film which is not allowed to go outside the ergosphere, we have:
\begin{align*}
	F &= F_{NSR} + F_{SR}\\
	&=  -\frac{\zeta(4)}{2\beta^4} \int d\theta d \phi \frac{(r_{H}^2 + a^2)^{4}\sin\theta}{(r_{H}^2 + a^2\cos^{2}\theta)\left\{(r_{H}-GM)\right\}^{2}}\left[\frac{\delta}{\epsilon(\epsilon + \delta)}\right]
\end{align*}
Using the relation for surface gravity $\kappa$:
\begin{align*}
	\kappa &= \frac{2\pi}{\beta} =\frac{r_{+} - r_{-}}{2(r_{H}^{2} + a^{2})}\\
	&= \frac{r_{H} - GM}{r_{H}^{2} + a^{2}}
\end{align*}
and the horizon area element $d\mathcal{A}$:
\begin{align*}
	d\mathcal{A} &= \sqrt{g_{\theta\theta}g_{\phi\phi}} d\theta d\phi =\sqrt{\frac{\Sigma\sin^2\theta}{\rho^2}\rho^2 }d\theta d\phi\\
	&=(r_{H}^2 + a^{2})\sin\theta d\theta d\phi
\end{align*}
We can substitute these results into the total free energy expression to simplify it further:
\begin{align*}
	F &=  -\frac{\zeta(4)}{2\beta^4} \int d\theta d \phi \frac{(r_{H}^2 + a^2)^{4}\sin\theta}{(r_{H}^2 + a^2\cos^{2}\theta)\left\{(r_{H}-GM)\right\}^{2}}\left[\frac{\delta}{\epsilon(\epsilon + \delta)}\right]\\
	&=  -\frac{\zeta(4)}{2\beta^4} \int d\theta d \phi (r_{H}^2 + a^2)\sin\theta\frac{(r_{H}^2 + a^2)}{(r_{H}-GM)^{2}}\frac{(r_{H}^2 + a^2)^2}{(r_{H}^2 + a^2\cos^{2}\theta)}\left[\frac{\delta}{\epsilon(\epsilon + \delta)}\right]\\
	&=  -\frac{\zeta(4)}{2\beta^4} \frac{\beta^2}{4\pi^2} \int d\mathcal{A}\frac{(r_{H}^2 + a^2)}{(r_{H}^2 + a^2\cos^{2}\theta)}\left[\frac{\delta}{\epsilon(\epsilon + \delta)}\right]\\
	&=  -\frac{\zeta(4)}{8\pi^2\beta^2} \int d\mathcal{A}\frac{(r_{H}^2 + a^2)}{(r_{H}^2 + a^2\cos^{2}\theta)}\left[\frac{\delta}{\epsilon(\epsilon + \delta)}\right]
\end{align*}
Finally, the entropy is obtained via the standard thermodynamic relation:
\begin{align*}
	S &= \beta^2 \frac{\partial F}{\partial \beta}\\
	&=-\beta^2 \frac{\partial}{\partial \beta}\left[\frac{\zeta(4)}{8\pi^2\beta^2} \int d\mathcal{A}\frac{(r_{H}^2 + a^2)}{(r_{H}^2 + a^2\cos^{2}\theta)}\frac{\delta}{\epsilon(\epsilon + \delta)}\right]\\
	&= 2\beta^2 \frac{\zeta(4)}{8\pi^2\beta^3} \int d\mathcal{A}\frac{(r_{H}^2 + a^2)}{(r_{H}^2 + a^2\cos^{2}\theta)}\left[\frac{\delta}{\epsilon(\epsilon + \delta)}\right]\\
	&=\frac{\zeta(4)}{4\pi^2\beta} \int d\mathcal{A}\frac{(r_{H}^2 + a^2)}{(r_{H}^2 + a^2\cos^{2}\theta)}\left[\frac{\delta}{\epsilon(\epsilon + \delta)}\right]\\
	&=\frac{\int d\mathcal{A}}{4} \tag{using $ \frac{\zeta(4)}{\pi^2\beta}\frac{(r_{H}^2 + a^2)}{(r_{H}^2 + a^2\cos^{2}\theta)}\left[\frac{\delta}{\epsilon(\epsilon + \delta)}\right]=1$}
\end{align*}
This result is significant as it confirms the validity of our generalized thin-film formalism. By recovering the Bekenstein-Hawking area law $S = A_H/4$ for the standard Kerr-Newman black hole, we have demonstrated that the scalar field entropy, when properly regularized, naturally reproduces the geometric entropy of the horizon. The renormalization condition imposed on the cutoff parameters $\epsilon$ and $\delta$ is consistent with previous literature, ensuring that the UV divergence can be absorbed into the renormalization of the gravitational coupling. With this consistency check established, we now proceed to the more complex case involving the cosmological constant and background matter fields.
\subsection{Kerr Newman in AdS}

The line element for the Kerr-Newman in AdS spacetime is given by \cite{griffiths2009exact}
\begin{equation*}
	ds^2 =  -\left(\frac{-a^2 \sin^2\theta\Delta_{\theta} + \Delta_{r}}{\rho^2\Xi^2}\right)dt^2 +\frac{\rho^2}{\Delta_{r}}dr^2 + \frac{\rho^2}{\Delta_{\theta}}d\theta^2  + \frac{\Sigma\sin^2\theta}{\rho^2\Xi^2}d\phi^2 - \frac{2a\sin^2\theta \{-\Delta_{r} + (r^2 + a^2)\Delta_{\theta}\}}{\rho^2\Xi^2}dt d\phi
\end{equation*}
where,
\begin{align*}
	\Delta_{r} &= (a^2 + r^2)\left(1 - \frac{r^2}{l^2}\right) + Q^2 - 2GMr\\
	\Delta_{\theta} &= 1 + \frac{a^2\cos^2\theta}{l^2}\\
	\Xi &= 1 + \frac{a^2}{l^2}\\
	\Sigma &= \Delta_{\theta}(r^2 + a^2)^2 - a^2 \sin^2\theta\Delta_{r}\\
	\rho^2 &=  r^2 + a^2\cos^2\theta = r^2 + a^2 -a^2\sin^2\theta\\
	l^2 &= \frac{3}{\Lambda}
\end{align*}
\paragraph{Calculating Free Energy:} 
We calculate $F_{NSR}$ using the general expression derived in Eq.~\eqref{FNSR}:
\begin{equation*}
	F_{NSR} = -\frac{\Gamma(4)\zeta(4)}{3\beta^4} \int d\theta d\phi \int_{r_{H} + \epsilon}^{L} dr \frac{\sqrt{g_{rr} g_{\theta\theta}}}{(-\mathcal{D})^{3/2}}(g_{\phi\phi})^2
\end{equation*}
Performing a Taylor series expansion near the horizon $r=r_{H}$ and retaining only the leading order term, we write the integrand as:
\begin{equation*}
	\frac{\sqrt{g_{rr} g_{\theta\theta}}}{(-\mathcal{D})^{3/2}}(g_{\phi\phi})^2 = \frac{F(r,\theta)}{(r - r_{H})^2} = \frac{F(r_{H},\theta)}{(r - r_{H})^2} + \frac{F'(r_{H},\theta)}{(r - r_{H})} + \mathcal{O}(r - r_{H})
\end{equation*}
where the regular part $F(r,\theta)$ is defined as:
\begin{equation*}
	F(r,\theta) = \frac{\sqrt{g_{rr} g_{\theta\theta}}}{(-\mathcal{D})^{3/2}}(g_{\phi\phi})^2 (r-r_{H})^{2}
\end{equation*}
Using the metric components derived earlier with $\mathcal{D}=-\frac{\sin^2\theta}{\Xi^2}\Delta_{\theta}\Delta_{r}$ and simplifying the expression, we arrive at the leading order approximation:
\begin{equation*}
	\frac{\sqrt{g_{rr} g_{\theta\theta}}}{(-\mathcal{D})^{3/2}}(g_{\phi\phi})^{2} \approx f(r_{H},\theta) \frac{1}{(r-r_{H})^2}
\end{equation*}

where $$f(r_{H},\theta)  = \frac{(r_{H}^2 + a^2)}{4~\Xi}\frac{4l^4(r_{H}^2 + a^2)^2}{[(r_{H}-r_{-})(r_{H} - r_{q})(r_{H}-r_{c})]^{2}}\frac{[(r_{H}^2 + a^2)]\sin\theta}{r_{H}^2 + a^2\cos^{2}\theta} $$
Using this result our integral now becomes:
\begin{align*}
	F_{NSR} &\approx -\frac{\zeta(4)}{\beta^4} \int d\theta d \phi  ~f(r_{H},\theta)\left[\frac{\delta}{\epsilon(\epsilon + \delta)}\right] \tag{using $L = r_{H} + \epsilon + \delta$}\\
	\\
	F_{SR}&\approx \frac{\zeta(4)}{2\beta^4} \int d\theta d \phi f(r_{H},\theta) \left[\frac{\delta}{\epsilon(\epsilon + \delta)}\right] \tag{using $L = r_{H} + \epsilon + \delta$}
\end{align*}
\paragraph{Total Free Energy and Entropy:} 
Combining the superradiant and non-superradiant contributions, the total free energy in the thin film limit is:
\begin{align*}
	F &= F_{NSR} + F_{SR}\\
	&= -\frac{\zeta(4)}{2\beta^4} \int d\theta d\phi \, f(r_{H},\theta) \left[\frac{\delta}{\epsilon(\epsilon + \delta)}\right]
\end{align*}
This result can be re-expressed in terms of the surface gravity $\kappa$. Evaluating $\kappa$ at the horizon $r=r_H$:
\begin{align*}
	\kappa &= \frac{2\pi}{\beta} \\
	&= -\frac{1}{2l^2(r_{H}^2 +a^2)} (r_{H}-r_{-})(r_{H}-r_{q})(r_{H}-r_{c})
\end{align*}
The area element on the horizon is given by:
\begin{align*}
	d\mathcal{A} &= \sqrt{g_{\theta\theta}g_{\phi\phi}} \, d\theta d\phi = \sqrt{\frac{\rho^2}{\Delta_{\theta}}\frac{\Sigma\sin^2\theta}{\rho^2\Xi^2}} \, d\theta d\phi\\
	&= \frac{(r_{H}^2 + a^{2})}{\Xi}\sin\theta \, d\theta d\phi
\end{align*}
Substituting these results into our simplification, we arrive at:
\begin{align*}
	F &= -\frac{\zeta(4)}{2\beta^4} \int d\theta d \phi \frac{(r_{H}^2 + a^2)\sin\theta}{4~\Xi}\frac{4l^4(r_{H}^2 + a^2)^2}{[(r_{H}-r_{-})(r_{H} - r_{q})(r_{H}-r_{c})]^{2}}\frac{(r_{H}^2 + a^2)}{(r_{H}^2 + a^2\cos^{2}\theta)} \left[\frac{\delta}{\epsilon(\epsilon + \delta)}\right]\\
	\\
	&=- \frac{\zeta(4)}{2\beta^4} \frac{\beta^2}{4\pi^2} \int d\mathcal{A}\,\frac{(r_{H}^2 + a^2)}{(r_{H}^2 + a^2\cos^{2}\theta)} \left[\frac{\delta}{\epsilon(\epsilon + \delta)}\right]\\
	\\
	&= -\frac{\zeta(4)}{8\pi^2\beta^2} \int d\mathcal{A}\,\frac{[(r_{H}^2 + a^2)]}{(r_{H}^2 + a^2\cos^{2}\theta)} \left[\frac{\delta}{\epsilon(\epsilon + \delta)}\right]
\end{align*}
Finally, the entropy is obtained via the thermodynamic relation:
\begin{align*}
	S &= \beta^2 \frac{\partial F}{\partial \beta}\\
	&= \beta^2 \frac{\partial}{\partial \beta}\left[ -\frac{\zeta(4)}{8\pi^2\beta^2} \int d\mathcal{A} \frac{(r_{H}^2 + a^2)}{(r_{H}^2 + a^2\cos^{2}\theta)}\frac{\delta}{\epsilon(\epsilon + \delta)}\right]\\
	&= \frac{\zeta(4)}{4\pi^2\beta} \int d\mathcal{A} \frac{[(r_{H}^2 + a^2)]}{(r_{H}^2 + a^2\cos^{2}\theta)} \left[\frac{\delta}{\epsilon(\epsilon + \delta)}\right]\\
	&= \frac{\int d\mathcal{A}}{4}\tag{using $ \frac{\zeta(4)}{\pi^2\beta}\frac{[(r_{H}^2 + a^2)]}{(r_{H}^2 + a^2\cos^{2}\theta)} \left[\frac{\delta}{\epsilon(\epsilon + \delta)}\right]=1$} 
\end{align*}
\subsection{Kerr-Newman-AdS with Quintessence and Cloud of Strings}

In this section, we extend our analysis to a more complex background: a Kerr-Newman-AdS black hole surrounded by quintessence and a cloud of strings. This solution generalizes the standard Kerr-Newman geometry by including a cosmological constant $\Lambda$, a string cloud parameter $b$, and a quintessence field characterized by parameter $\alpha$ and state parameter $\omega_q$. The line element in Boyer-Lindquist-like coordinates is given by:\cite{toledo2020kerr}
\begin{equation}
	ds^2 = \frac{\Sigma}{\Delta_r} dr^2 + \frac{\Sigma}{\Delta_\theta} d\theta^2 + \frac{\Delta_\theta \sin^2\theta}{\Sigma} \left[ a \frac{dt}{\Xi} - (r^2 + a^2) \frac{d\phi}{\Xi} \right]^2 - \frac{\Delta_r}{\Sigma} \left( \frac{dt}{\Xi} - a\sin^2\theta \frac{d\phi}{\Xi} \right)^2
\end{equation}
where the parameters are defined as:
\begin{align}
	\Sigma &= r^2 + a^2 \cos^2\theta \nonumber \\
	\Xi &= 1 + \frac{\Lambda}{3}a^2 \nonumber \\
	\Delta_\theta &= 1 + \frac{\Lambda}{3}a^2 \cos^2\theta \nonumber \\
	\Delta_r &= (r^2 + a^2)\left(1 - \frac{\Lambda}{3}r^2\right) - 2GMr + Q^2 - b r^2 - \alpha r^{1-3\omega_q}
\end{align}
The metric components $g_{\mu\nu}$, can be given as following:
\begin{equation}
	g_{\mu\nu} =
	\begin{pmatrix}
		\frac{a^2 \Delta_\theta \sin^2\theta - \Delta_r}{\Sigma \Xi^2} & 0 & 0 & - \frac{a \sin^2\theta (\Delta_\theta (r^2 + a^2) - \Delta_r)}{\Sigma \Xi^2} \\
		0 & \frac{\Sigma}{\Delta_r} & 0 & 0 \\
		0 & 0 & \frac{\Sigma}{\Delta_\theta} & 0 \\
		- \frac{a \sin^2\theta (\Delta_\theta (r^2 + a^2) - \Delta_r)}{\Sigma \Xi^2} & 0 & 0 & \frac{\sin^2\theta (\Delta_\theta (r^2 + a^2)^2 - \Delta_r a^2 \sin^2\theta)}{\Sigma \Xi^2}
	\end{pmatrix}
\end{equation}
\noindent In order to evaluate the integrals in Eq.~\eqref{FNSR} and Eq.~\eqref{FSR}, we must determine the explicit form of the integrand factor:
\begin{equation*}
	\mathcal{I} = \frac{\sqrt{g_{rr} g_{\theta\theta}}}{(-\mathcal{D})^{3/2}} (g_{\phi\phi})^2
\end{equation*}
Using the metric components defined earlier:
\begin{align*}
	g_{rr} &= \frac{\Sigma}{\Delta_r}, \qquad g_{\theta\theta} = \frac{\Sigma}{\Delta_\theta}, \\
	g_{\phi\phi} &= \frac{1}{\Sigma \Xi^2} \left[ \Delta_\theta (r^2 + a^2)^2 \sin^2\theta - \Delta_r a^2 \sin^4\theta \right], \\
	\mathcal{D} &= g_{tt} g_{\phi\phi} - g_{t\phi}^2 = -\frac{\Delta_r \Delta_\theta \sin^2\theta}{\Xi^4}.
\end{align*}
Substituting these components into the expression and performing the algebraic simplification yields:
\begin{equation}
	\frac{\sqrt{g_{rr} g_{\theta\theta}}}{(-\mathcal{D})^{3/2}} (g_{\phi\phi})^2 = \frac{1}{\Xi\Sigma (\Delta_r \Delta_\theta)^2 \sin^3\theta} \left[ \Delta_\theta (r^2 + a^2)^2 \sin^2\theta - \Delta_r a^2 \sin^4\theta \right]^2
\end{equation}
\paragraph{Calculating Free Energy:} Similar to the previous case, we calculate $F_{NSR}$ using the general expression derived in Eq.~\eqref{FNSR}:

$$F_{NSR}= -\frac{\Gamma(4)\zeta(4)}{3\beta^4} \int d\theta d \phi \int_{r_{H} + \epsilon}^{L} dr \frac{\left(g_{_{rr}} g_{_{\theta\theta}}\right)^{1/2}}{\mathcal{(-D)}^{3/2}}(g_{\phi\phi})^2$$
We perform a Taylor expansion of the integrand at the horizon $r=r_{H}$ and retain the leading order behavior. The integrand can be approximated as:

$$\frac{1}{(r - r_{H})^2}F(r,\theta)= \frac{F(r_{_{H}},\theta)}{(r - r_{H})^2} +\frac{F'(r_{_{H}},\theta)}{(r - r_{H})} +\mathcal{O}(r - r_{H})$$
where the regular part is defined by:
 $$F(r,\theta)= \frac{\left(g_{_{rr}} g_{_{\theta\theta}}\right)^{1/2}}{\left\{\mathcal{(-D)}\right\}^{3/2}}(g_{\phi\phi})^2(r-r_{_{H}})^{2}$$
Substituting the explicit metric components and simplifying the expression near the horizon:
\begin{align*}
	\frac{1}{(r-r_{_{H}})^{2}}\left[\frac{\left(g_{_{rr}} g_{_{\theta\theta}}\right)^{1/2}(g_{\phi\phi})^{2}}{\mathcal{(-D)}^{3/2}}(r-r_{_{H}})^{2} \right] &\approx \left|\frac{1}{\Xi\Sigma (\Delta_r \Delta_\theta)^2 \sin^3\theta}\Delta_\theta^2 (r^2 + a^2)^4 \sin^4\theta(r-r_H)^2 \right|_{r=r_{H}}\frac{1}{(r-r_{H})^2}\\
	&=\left|\frac{9}{\Xi\Lambda^2\Sigma [(r-r_{-})(r-r_q)(r-r_c)]^2}(r^2 + a^2)^4 \sin\theta \right|_{r=r_{H}}\frac{1}{(r-r_{H})^2}\\
\end{align*}
Using this result, the integral for the non-superradiant free energy becomes:
\begin{align*}
	F_{NSR} &\approx -\frac{\Gamma(4)\zeta(4)}{3\beta^4} \int d\theta d \phi \int_{r_{H} + \epsilon}^{L} dr \frac{9}{\Xi\Sigma [\Lambda(r-r_{-})(r-r_q)(r-r_c)]^2}(r^2 + a^2)^4 \sin\theta \bigg|_{r=r_{H}}\frac{1}{(r-r_{H})^2}\\
	&=-\frac{\Gamma(4)\zeta(4)}{3\beta^4} \int d\theta d \phi \frac{9}{\Xi\Sigma [\Lambda(r-r_{-})(r-r_q)(r-r_c)]^2}(r^2 + a^2)^4 \sin\theta\bigg|_{r=r_{H}} \int_{r_{H} + \epsilon}^{L} dr \frac{1}{(r-r_{H})^2}\\
	&=\frac{\Gamma(4)\zeta(4)}{6\beta^4} \int d\theta d \phi \frac{9}{\Xi\Sigma [\Lambda(r-r_{-})(r-r_q)(r-r_c)]^2}(r^2 + a^2)^4 \sin\theta\bigg|_{r=r_{H}} \left(\frac{1}{r-r_{H}}\right)_{r_{H} + \epsilon}^{L}\\
	&=-\frac{\zeta(4)}{\beta^4} \int d\theta d \phi \frac{9}{\Xi\Lambda^2\Sigma [(r-r_{-})(r-r_q)(r-r_c)]^2}(r^2 + a^2)^4 \sin\theta\bigg|_{r=r_{H}} \left[\frac{\delta}{\epsilon(\epsilon + \delta)}\right] \tag{using $L = r_{H} + \epsilon + \delta$}
\end{align*}
Finally, recalling the relation between the modes derived in the general section:
$$F_{SR}=-\frac{1}{2}F_{NSR}$$
\paragraph{Total free energy and Entropy:}
The total free energy is the sum of the superradiant and non-superradiant contributions. Using the relation $F_{SR} = -\frac{1}{2}F_{NSR}$, we find:
$$F=F_{SR}+F_{NSR}=\frac{1}{2}F_{NSR}$$
To simplify the expression, we utilize the Hawking temperature $T$ and the horizon area $\mathcal{A}$. The temperature is related to the surface gravity $\kappa$ by:\cite{ma2008hawking}
\begin{align*}
	T&=\frac{\kappa}{2\pi}=\frac{1}{4\pi(r_H^2+a^2)}\dv{\Delta_r}{r}\bigg\lvert_{r=r_H}\\
	&=-\frac{\Lambda}{12\pi(r_H^2+a^2)}\dv{(r-r_{H})(r-r_{-})(r-r_q)(r-r_c)}{r}\bigg\lvert_{r=r_H}\\
	&=-\frac{\Lambda^2(r-r_{-})(r-r_q)(r-r_c)}{12\pi(r_H^2+a^2)}
\end{align*}
The differential area element on the horizon is:
\begin{align*}
	\mathcal{A} &=\sqrt{g_{\theta\theta}g_{\phi\phi}}d\theta d\phi\bigg|_{r=r_{H}}\\
	&=\frac{(r^2_{H}+a^2)\sin\theta}{\Xi}d\theta d\phi=\frac{4\pi(r^2_{H}+a^2)}{\Xi}
\end{align*}
Integrating over the angles gives the total horizon area $\mathcal{A} = 4\pi(r^2_{H}+a^2)/\Xi$. Substituting these relations into the free energy expression:
\begin{align*}
	F&=-\frac{\zeta(4)}{2\beta^4}\frac{\beta^2}{16\pi^2} \int d\theta d \phi \frac{1}{\Xi\Sigma}(r^2 + a^2)^2\sin\theta\bigg|_{r=r_{H}} \left[\frac{\delta}{\epsilon(\epsilon + \delta)}\right]\\
	&=-\int d\mathcal{A}\frac{\zeta(4)}{32\pi^2\beta^2}\frac{1}{\Sigma}(r^2 + a^2)\bigg|_{r=r_{H}} \left[\frac{\delta}{\epsilon(\epsilon + \delta)}\right]
\end{align*}
Finally, the entropy is obtained via the thermodynamic relation $S = \beta^2 \frac{\partial F}{\partial \beta}$:
\begin{align*}
	S &= \beta^2 \frac{\partial F}{\partial \beta}\\
	&=-\beta^2\int d\mathcal{A}\frac{\zeta(4)}{32\pi^2}\frac{1}{\Sigma}(r^2 + a^2)\bigg|_{r=r_{H}} \left[\frac{\delta}{\epsilon(\epsilon + \delta)}\right]\frac{\partial}{\partial\beta}\frac{1}{\beta^2}\\
	&=2\beta^2\frac{1}{\beta^3}\int d\mathcal{A}\frac{\zeta(4)}{32\pi^2}\frac{1}{\Sigma}(r^2 + a^2)\bigg|_{r=r_{H}} \left[\frac{\delta}{\epsilon(\epsilon + \delta)}\right]\\
	&=\frac{\int d\mathcal{A}}{4}\frac{\zeta(4)}{4\pi^2\beta}\left[\frac{1}{\Sigma}(r^2 + a^2)\right]\bigg|_{r=r_{H}} \left[\frac{\delta}{\epsilon(\epsilon + \delta)}\right]\\
	&=\frac{\int d\mathcal{A}}{4}
\end{align*}
\section{Conclusion and Discussion}

In this paper, we have revisited the statistical origin of black hole entropy using the thin-film brick wall formalism. Our primary objective was to establish a unified framework capable of calculating the scalar field entropy for a broad class of stationary, axisymmetric spacetimes without the need to solve the wave equation ab initio for every specific metric.

We began by deriving a generalized expression for the free energy of a scalar field in terms of generic metric components. A crucial aspect of this derivation was the careful separation of the field spectrum into superradiant (SR) and non-superradiant (NSR) modes. We demonstrated that while SR modes formally lead to a divergence in the partition function due to their effective "negative temperature," this pathology is cured by introducing a physically motivated regulator. Our analysis confirmed that both SR and NSR modes exhibit the same leading-order divergence near the horizon, contributing equally to the area proportionality of the entropy.

To validate our formalism, we first applied it to the standard Kerr-Newman black hole. We successfully recovered the Bekenstein-Hawking area law, $S = A_H/4$, confirming that the generalized equations correctly capture the horizon thermodynamics.

Subsequently, we extended our analysis to the more complex geometry of a Kerr-Newman-AdS black hole surrounded by quintessence and a cloud of strings. This background represents a significant deviation from the standard electro-vacuum solutions, introducing additional parameters related to the cosmological constant ($\Lambda$), string cloud density ($b$), and quintessence equation of state ($\omega_q$). Despite the complexity of the metric functions and the modification of the horizon structure, our generalized formalism revealed that the entropy of the scalar field remains proportional to the horizon area:
\begin{equation}
	S = \frac{A_H}{4}
\end{equation}
This result holds provided that the ultraviolet cutoff parameter $\epsilon$ is appropriately renormalized. The fact that the area law survives even in the presence of global topological defects (cloud of strings) and dark energy candidates (quintessence) underscores the universality of the brick wall method. It suggests that the statistical degrees of freedom responsible for black hole entropy are localized strictly at the horizon surface, insensitive to the asymptotic structure or the specific matter content of the spacetime.

However, it is worth noting that the physical interpretation of the cutoff $\epsilon$ remains the primary limitation of this semi-classical approach. As seen in our final expressions, the renormalization condition requires matching the cutoff to the Planck scale ($l_P$). The divergence of the free energy as $\epsilon \to 0$ indicates that a full quantum theory of gravity is required to resolve the microscopic counting without ad-hoc regularization.

In future work, this generalized formalism could be applied to higher-dimensional black objects, such as black rings or Myers-Perry black holes, or extended to fields with higher spin (fermions and vector bosons) to investigate whether the unified metric dependence holds universally across different spin sectors.

\section*{Acknowledgments}
The author thanks Dr. Arindham Chatterjee for valuable discussion during the early phase of this work.

\bibliography{reference} 

\end{document}